\documentclass[twocolumn, pra, showpacs,superscriptaddress,floatfix]{revtex4-1}

\usepackage{eurosym}
\usepackage{amssymb}
\usepackage{mathrsfs}
\usepackage{graphicx}
\usepackage{float}
\usepackage{dcolumn}
\usepackage{bm}

\begin{document}

\title{Topologically nontrivial states in one-dimensional nonlinear bichromatic superlattices}
\author{C. S. Liu}
\affiliation{Department of Physics, Yanshan University, Qinhuangdao 066004,
China}

\author{Z. Z. Wang}
\affiliation{Department of Physics, Yanshan University, Qinhuangdao 066004,
China}

\author{Chuanhao Yin}
\affiliation{Beijing National Laboratory for Condensed Matter Physics, Institute of Physics, Chinese Academy of Sciences, Beijing 100190, China }

\author{Y. D. Wu}
\affiliation{Department of Physics, Yanshan University, Qinhuangdao 066004,
China}

\author{T. F. Xu}
\affiliation{Department of Physics, Yanshan University, Qinhuangdao 066004,
China}

\author{L. H. Wen}
\affiliation{Department of Physics, Yanshan University, Qinhuangdao 066004,
China}

\author{Shu Chen}
\thanks{Corresponding author, schen@iphy.ac.cn}
\affiliation{Beijing National Laboratory for Condensed Matter Physics, Institute of Physics, Chinese Academy of Sciences, Beijing 100190, China }
\affiliation{Collaborative Innovation Center of Quantum Matter, Beijing, China}
\date{ \today }

\begin{abstract}
We study topological properties of one-dimensional nonlinear bichromatic superlattices and unveil the effect of nonlinearity on topological states. We find the existence of nontrivial edge solitions, which distribute on the boundaries of the lattice with their chemical potential located in the linear gap regime and are sensitive to the phase parameter of the superlattice potential. We further demonstrate that the topological property of the nonlinear Bloch bands can be characterized by topological Chern numbers defined in the extended two-dimensional parameter space. In addition, we discuss that the composition relations between the nolinear Bloch waves and gap solitions for the nonlinear superlattices. The stabilities of edge solitons are also studied.
\end{abstract}

\pacs{03.75.Lm, 05.30.Jp, 73.21.Cd }
\maketitle






\section{Introduction}

In recent years, there are growing efforts in studying  one-dimensional (1D) periodic and quasiperiodic superlattice systems with nontrivial topological properties, which can be experimentally realized in cold atomic systems and photonic systems \cite{Atala,Nature.453.895,
NaturePhys.6.354,Kraus}.
The 1D optical superlattice can be produced by superimposing two 1D optical lattices with
different wavelengths \cite{Nature.453.895,
NaturePhys.6.354}. Recently, it was shown that the 1D optical superlattice systems can exhibit rich topological phases \cite{Lang,Grusdt,Zhu,Deng,Guo,Xu}. 
Additionally, nontrivial topological edge states in 1D photonic quasicrystals
have also been observed experimentally \cite{Kraus}.  
As the topological properties of the 1D noninteracting superlattice systems can be understood from their band structures, it is interesting to study the interaction effect on the edge states of the bosonic superlattice systems, particularly, for the weakly interacting bosonic system in which the effect of interactions between bosons can be effectively described by nonlinear Schr\"{o}dinger equation.

In the scheme of mean field theory, it is well known that the interactions
between bosons can result in the significant nonlinearity in a periodic Bose system.
In the presence of both the nonlinearity and periodicity, there exist two kinds of
important waves, namely nonlinear Bloch waves and gap solitons.
While Bloch waves are intrinsic to periodic systems and are extensive
over the whole space, nonlinearity has a significant influence on their stabilities \cite{Aschcoft}. The instabilities are responsible for the formation of the train
of localized filaments \cite{PhysRevLett.92.163902} and are closely related to the
breakdown of superfluidity \cite{PhysRevLett.86.4447}.
Gap solitons are spatially localized wave
packets with the chemical potentials in the linear band gaps \cite{book}.
According to the locations of their chemical potential, gap solitons can
be divided into several classes. For example, when the chemical potentials are in the
linear band gaps, the localized wave packets have a major peak well
localized within a unit cell, and are called  the fundamental gap
soliton \cite{PhysRevA.67.013602}.
It has been shown that there exists a composition relation between them: Bloch waves at either the center
or edge of the Brillouin zone are infinite chains composed of fundamental gap solitons \cite{PhysRevLett.102.093905}.
In this work, we shall study the effect of nonlinearity on the topological properties of 1D bichromatic superlattices. The interesting questions include whether the topological states in the noninteracting limit can survive in the presence of nonlinearity, and whether the gap solitons can be formed in the bichromatic superlattice systems? If the gap solitions exsit, what are their relations to the topological states and whether a composition relation between the nolinear Bloch waves and gap solitions still holds ture?

To answer these questions, we study the interacting boson system trapped in a 1D optical superlattice, which is described by a nonlinear Schr\"{o}dinger equation in a 1D bichromatic periodic potential. 
By numerically solving the nonlinear Schr\"{o}dinger equation under the periodic boundary condition, we find the existence of nonlinear Bloch waves, which form a nonlinear Bloch band adiabatically connected to the topological Bloch band in the noninteracting limit. For the system under the open boundary condition, we find the existence of edge gap solitons and discuss their stabilities. The edge gap soliton can be viewed as a reminiscence of the topologically nontrivial edge state for the noninteracting bichromatic superlattice. We verify the existence of a series of gap solitons for the system under the periodic boundary condition,
and the composition relations between the nolinear Bloch waves and gap solitions are also discussed.
The paper is organized as follows. In Sec. \ref{Theory}, we introduce the theoretical model and show how the 1D bichromatic superlattice system can be mapped to the Harper-Hofstadter problem. In Sec. \ref{Results}, we first present the spectrum of the nonlinear superlattice system in the subsection A. The edge states and the topological properties of the nonlinear Bloch band are discussed in the subsection B. The composition relations between gap solitons and nonlinear Bloch waves, are investigated in the subsection C. The stabilities of edge solitons are discussed in the subsection D. Sec.~\ref{Summary} gives a brief summary.

\section{Model}

\label{Theory}

We consider a weakly interacting Bose gas loaded in 1D  optical
superlattice confined in $\left[ -L/2,L/2\right] $. On the mean field level, the
above system can be well described by the following nonlinear Schr\"{o}%
dinger equation
\begin{equation}
\left[ -\frac{\hbar ^{2}}{2m}\frac{d^{2}}{dx^{2}}+V(x)+g\left\vert \Psi
(x)\right\vert ^{2}\right] \Psi (x)=\mu _{n}\Psi (x),
\label{nonlinear Schrodinger equation}
\end{equation}%
where $m$ is the mass of bosons, $\mu _{n}$ is the chemical potential which adiabatically connects to the $n$-th single
particle eigenvalue when $g \rightarrow 0$, the wave function is normalized under $%
\int_{-L/2}^{L/2}\left\vert \Psi (x)\right\vert ^{2}dx=1$, and $g$ is the
effective interaction between bosons. The bichromatic periodic potential is given by
\begin{eqnarray}
V(x) &=&V_{1}(x)+V_{2}(x)  \nonumber \\
&=&v_{1}\mathrm{cos}(2\pi x)+v_{2}\mathrm{cos}(2\pi \alpha x+\delta ),
\label{quasi-periodic potential}
\end{eqnarray}%
where $v_{1}$ and $v_{2}$ are the potential strength, $\alpha $ is a rational
number, and $\delta $ is an arbitrary phase. The bichromatic superlattices have been realized in cold atomic experiments \cite{Atala,Nature.453.895,
NaturePhys.6.354}. Besides, the nonlinear periodic
systems can also be realized in nonlinear waveguide arrays \cite%
{Nature.424.817, PhysRevLett.81.3383} and optically induced lattices \cite%
{Nature.422.147}.

Despite the existence of nonlinearity, Eq. (\ref{nonlinear Schrodinger equation}) under
the periodic boundary condition still has the Bloch wave solutions $\Psi
\left( x\right) =e^{ikx}\psi _{k}\left( x\right) $, where $k$ is the Bloch
wave vector.  For the system with $\alpha =1/q$ and $q$ being a positive integer, the Bloch wave state $\psi _{k}\left( x\right) $ is a periodic
function, which fulfills $\psi _{k}\left( x\right) =\psi _{k}\left( x+a\right) $ with $a=1/\alpha$
being the period of potential function $V\left( x\right) $. From the Schr\"{o}dinger equation (\ref{nonlinear Schrodinger equation}), we have the
following equation for each Bloch wave state $\psi _{k}\left( x\right) $
\begin{eqnarray}
&&\left[ -\frac{\hbar ^{2}}{2m}\left( \frac{d}{dx}+ik\right) ^{2}+V\left(
x\right) +g\left\vert \psi _{k}\left( x\right) \right\vert ^{2}\right] \psi
_{k}\left( x\right)  \nonumber \\
&=&\mu _{kn}\psi _{k}\left( x\right) .  \label{Bloch equation}
\end{eqnarray}%
However, under the open boundary condition, the momentum $k$ is no longer a
good quantum number.

As there are no analytic solutions for the above two nonlinear equations [Eqs. (\ref{nonlinear Schrodinger equation}) and (\ref{Bloch equation})], several numerical methods have been used to solve them \cite{Wu2}. A very practical method we used in the present work is as the following. The equations are first solved by finite difference method in the linear case ($g=0$) to obtain the eigenvalue and eigenstate. Then the eigenstate is brought back to the equation with the effective potential function $V\left( x\right) +g\left\vert \Psi\left( x\right) \right\vert ^{2}$ and get the new the eigenvalue and eigenstate. Iterating the above step several times, we can find the stable eigenvalue and eigenstate. For the nonlinear Schr\"{o}dinger equation (\ref{nonlinear Schrodinger equation}), the different interval is taken to $[-L/2, L/2]$, where $L$ is the region of periodic potential $V(x)$. For the nonlinear Bloch equation (\ref{Bloch equation}), the different interval is taken to $[0,a]$.

When $v_{2}$ is much smaller than $v_{1}$, the potential $V_{2}(x)$ can be
taken as a perturbation in Eq. (\ref{nonlinear Schrodinger equation}). In the
case of the large potential strength $v_{1}$, the low-energy orbitals are
localized in the unit cell of the periodic potential $V_{1}(x)$. Their
hopping integrals involving second or further apart neighbors are negligible. The above model can be effectively described by
a tight-binding model with periodic on-site potentials \cite{Modugno}
\begin{eqnarray}
-J (u_{i+1}+u_{i-1})
+ \Delta \mathrm{cos}(2\pi \alpha i + \delta ) u_{i} \nonumber \\
+ c \left\vert u_{i}\right\vert ^{2}u_{i}
= \epsilon_{n} u_{i},  \label{Harper}
\end{eqnarray}%
where $u_{i}$ is the amplitude of the particle wave function at the $i$-th
site and $J$ is the hopping integral of the nearest neighbors, $\Delta \propto v_{2}$, and $c \propto g$. In the noninteracting limit of  $c=0$, the tight-binding model [Eq. (\ref{Harper})] reduces to the well-known Aubry-Andr\'{e} (AA) model \cite{Ann.Isr.Phys.Soc.3.133} or Harper-Hofstadter model \cite{Proc.Phys.Soc.A68.874,PhysRev.14.2239}. The topological properties of the AA model have been unveiled in Ref. \cite{Lang,Kraus} by demonstrating the existence of nontrivial edge states and topological invariants in the two-dimensional parameter space through dimensional extension.

\section{Results and discussions}

\label{Results}

\subsection{Energy spectrum}

\begin{figure}[tbp]
\includegraphics[width=9cm]{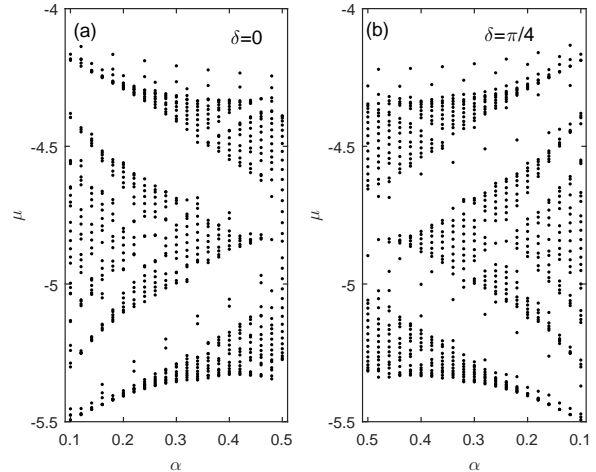}
\caption{The butterflylike energy spectra with respect to $\protect\alpha$
varying from $0.1$ to $0.5$ with different phases: (a) $\protect\delta =0$,
(b) $\protect\delta =\protect\pi /4$. Both figures are for the system with $%
v_1=16$, $v_2=0.02v_1$, $g=12$ and $L=34$ under the open boundary conditions.}
\label{Fig1}
\end{figure}

In the noninteracting tight-binding limit, it is known that the spectrum of the superlattice system for various $\alpha$ exhibits the butterfly structure as the system described by Eq. (\ref{Harper}) with $c=0$ can be mapped to the Hofstadter model \cite{PhysRev.14.2239,Lang}. To see how the structure of energy spectrum of the superlattice system is affected by the nonlinear term, we numerically solve the the nonlinear Schr\"{o}dinger equation (\ref{nonlinear Schrodinger equation}) under the open boundary condition and
plot the energy spectrum of Eq. (\ref{nonlinear Schrodinger equation}) versus different $\alpha$ with the interacting parameter $g=12$ in Fig. \ref{Fig1}. The other parameters are taken to be $v_{1}=16$, $v_{2}=0.02v_{1}$ and $L=34$, and the natural unit is used, i.e. $\hbar=m=1$.
We shall keep this set of parameters fixed in the following discussion. The energy spectrums of the 1D nonlinear superlattice system shows the similar butterfly structure as the spectrum of the noninteracting 1D superlattice system \cite{Lang}. The basic structures shown in Fig. \ref{Fig1} (a) and (b), corresponding to different phases $\delta =0$ and $\delta =\pi /4$, are quite similar. In the band gap regions of the butterfly structure, there are some isolated points which are corresponding to the edge states.
The position of the edge state is dependent on the value of $\delta$. 
\begin{figure}[tbp]
\includegraphics[width=9cm]{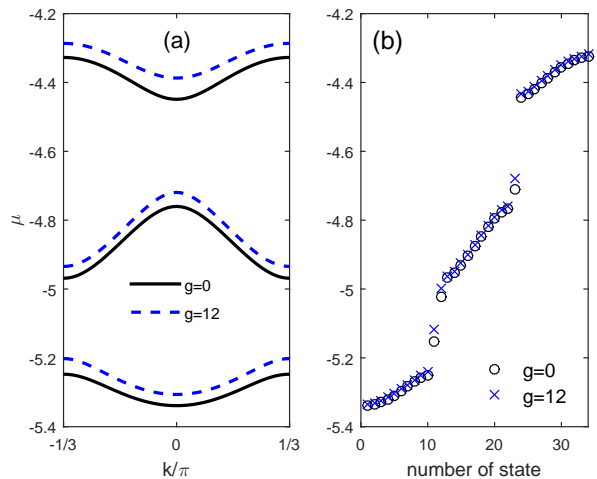}
\caption{Frame (a) compares the linear Bloch energy bands [the black solid
lines] with the nonlinear Bloch energy  bands [the blue dashed lines] of nonlinear Bloch Schr\"{o}dinger equation (\protect\ref%
{Bloch equation}). Frame (b) shows that, under the open boundary condition, the
eigenvalues of nonlinear Schr\"{o}dinger equation (\protect\ref%
{nonlinear Schrodinger equation}) in ascending order also form three energy bands. Two edge states
appear in the first band gap and one edge state appears in the second band
gap. The parameter $\protect\delta$ is set to be $0.22
\protect\pi$. }
\label{Fig2}
\end{figure}

The energy spectrum of the nonlinear superlattice system displays similar structure as the corresponding noninteracting system \cite{Lang}. To see it clearly, we consider the 3-period superlattice system and solve the nonlinear equations [Eqs. (\ref{nonlinear Schrodinger equation}) and (\ref{Bloch equation})] with $\alpha=1/3$,  $\delta =0.22\pi $ and $g=12$. The energy spectrum for the Bloch equation [Eq. (\ref{Bloch equation})] is shown in Fig. \ref{Fig2}(a). In the presence of the nonlinear term, the nonlinearity lifts the Bloch bands into gap regions of linear bands. When $g$ decreased to zero, the nonlinear bands move down continuously to their noninteracting limit. For the system under open boundary condition, the corresponding eigenvalues of nonlinear Schr\"{o}dinger equation (\protect\ref{nonlinear Schrodinger equation}) in ascending order also form three energy bands shown in Fig. \ref{Fig2}(b). The nonlinear bands marked by 'cross' originate from the linear band marked by 'circle'.

\subsection{Edge Solitons and Topological invariant}

\label{Edge Solitons and Topological invariant}

\begin{figure}[tbp]
\includegraphics[width=9cm]{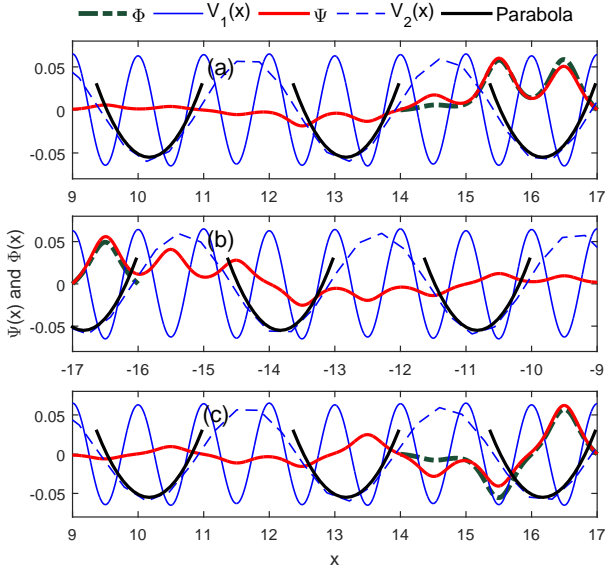}
\caption{The figures show the three edge states and explain how these states
can be induced by the parabolas and the walls. The $y$ coordinates show the amplitudes of the wave functions $\Psi(x)$ and $\Phi(x)$. In order to show the periodic potential $V_1(x)$ and $V_2(x)$ in this figure, the amplitudes of $V_1(x)$ and $V_2(x)$ are
been multiplied two rates 0.00375 and 0.15 respectively.
From top to bottom, the red
thick solid lines represent three edge
states of the nonlinear Schr\"{o}dinger equation (\ref{nonlinear
Schrodinger equation}) [(a): $\Psi_{11}(x)$, (b): $\Psi_{12}(x)$ and (c): $\Psi_{23}(x)$]. The green
dashed lines show the wave functions $\Phi(x)$ of the nonlinear Schr\"{o}dinger equation (\ref{nonlinear
Schrodinger equation}) under the parabolic approximation. The periodic
potential $V_1(x)$ are shown by the blue solid lines. The blue dashed lines
are used to present the 2th-periodic potential $V_2(x)$. The 2th-periodic potential $V_2(x)$ forms
several parabolas. The black solid lines show the parabolas. }
\label{Fig3}
\end{figure}

As shown in Fig. \ref{Fig2}, it is interesting to see that two states $\Psi _{11}$, $\Psi _{12}$ appear in the first nonlinear band gap and the state $\Psi _{23}$ appears in the second nonlinear band gap, where we have used the subscript $n$ to represent the $n$-th eigenstate $\Psi_{n}$ in ascending order.  Similar to the non-interaction case, these three states are edge states with the wave functions localized at the left or right boundaries,  as shown in Fig. \ref{Fig3} (marked by red thick solid lines). The formations of these edge states are
due to the interplay between the kinetic energy, the nonlinear interaction and the confined periodic potential. For convenience, we call these edge state as edge solitons.

To understand the origin of the edge solitons in nonlinear bichromatic superlattices in an intuitive way, we plot
the periodic potentials $V_{1}(x)$ (blue solid lines) and $V_{2}(x)$ (blue
dashed lines) in Fig. \ref{Fig3}. Since we are interested in the low energy states, the bottom of $V_2(x)$ is important.
Near the bottom of $V_2(x)$, the potential of unit cell
can be approximated by the parabolic potential. Considering the periodicity of $V_{2}(x)$, we obtain a serial of parabolas shown in Fig. \ref{Fig3} by black solid lines. Under the parabolic approximation, the Schr\"{o}dinger equation (\ref{nonlinear
Schrodinger equation}) remains unchanged when we shift the vertex of a parabola at $x_0$ a period to $x_0 \pm a$, if the vertex is away from the boundary. When the interacting bosons are confined in a parabola, the particles are in a series of the discrete eigenstates. A substitution of one parabola with the other parabola only shifts the parabola and wave function, and does not change its energy dramatically. This holds true until the parabola touches the boundary. In such case, the walls and the parabola provide the main confinement. The particles now sit in a roughly triangular potential well. Due to the stronger confinement, the energy levels will be elevated and higher than the corresponding levels in the middle. The corresponding states are squeezed against the side of the wall and the degeneracies are lost.

Under the parabolic approximation, we solve the
nonlinear Schr\"{o}dinger equation [Eq. (\ref{nonlinear Schrodinger equation})] and obtain the orbital wave functions of edge parabola shown in Fig. \ref{Fig3}
(green dashed lines). Comparing the red solid lines and the green dashed lines in Fig. \ref{Fig3}, the orbital wave functions are found to coincide with the edge solitons well.
For the three edge gap solitons, the wave functions localize on the boundary and trail a long tail.
For the former two edge gap solitons, i.e., $\Psi _{11}$ in Fig. \ref{Fig3} (a) and $\Psi _{12}$ in Fig. \ref{Fig3} (b), their chemical potentials are in the first band gap, and the corresponding wave functions develop from the ground states of the right and left edge parabola, respectively. The right edge parabola of the $\Psi _{11}$ includes two lattice of $V_{1}(x)$. So the wave function shows two peaks with the same sign. However, the left edge parabola of the $\Psi _{12}$ includes one lattice of $V_{1}(x)$. So the wave function shows only one peak. The left edge parabola for $\Psi _{12}$ is closer to the left wall than the right edge parabola for $\Psi _{11}$ in the right side, it gives a strong confinement of the particles. So the chemical potential of $\Psi_{12}$ is higher than that of $\Psi _{11}$. For the edge gap solitons $\Psi _{23}$ in Fig. \ref{Fig3} (c), the parabola of this state is the same as that of $\Psi_{11}$ in Fig. \ref{Fig3} (a). However their chemical potential is in the second band gap and the state develops from the first excited state. So the wave function also shows two peaks. One is positive and the other one is negative.

\begin{figure}[tbp]
\includegraphics[width=9cm]{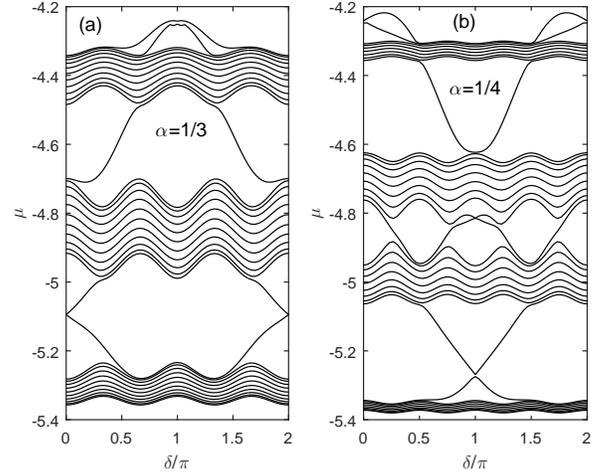}
\caption{Nonlinear energies vary with the phase $\protect\delta$ for Eq. (\protect
\ref{nonlinear Schrodinger equation}) with the parameter (a) $%
\protect\alpha=1/3$, (b) $\protect\alpha=1/4$ under open boundary
conditions.  The interaction parameters are taken as $g=5$.}
\label{Fig4}
\end{figure}

As the phase $\delta $ changes from $0$ to $2\pi $, the spectrum of the
nonlinear Schr\"{o}dinger equation (\ref{nonlinear Schrodinger equation})
for a given $\alpha $ changes periodically. The position of the edge states
in the gaps also varies continuously with the change of the phase $\delta $. In Fig. %
\ref{Fig4}, we show the spectrum of the superlattice systems with $\alpha
=1/3$ and $1/4$ versus $\delta $ under the open boundary condition. The
shade parts correspond to the band regions and the lines between bands represent
the spectra of edge states. The
position of the edge states in the gaps varies continuously with the
change of $\delta $. In particular, the level continuously connects the upper
and lower energy bands.

In general, the appearance of edge states is attributed to the nontrivial topological property of the bulk system, whose topological structure can be characterized by a topological invariant \cite{RevModPhys.82.3045, RevModPhys.83.1057}. To see this clearly for the problem considered in this work,
we also explore the topological properties of the nonlinear Bloch band under the periodic boundary condition. For the present nonlinear periodic system, the wave vector $k$ of nonlinear Bloch function can be changed form $0$ to $2\pi/q$ and the phase $\delta $ can also be varied from $0$ to $2\pi $ adiabatically, we therefore get a manifold of Hamiltonian $H(k,\delta )$ in the space $(k,\delta)$. An effective 2D Brillouin zone with respect to the Bloch vector $k$ and the potential shift $\delta $ forms a $T^{2}$ torus in the two directions. For eigenstates $\psi (k,\delta )$ of Bloch equation (\ref{Bloch equation}), the Chern number is used to characterize their topological properties. The Chern number is a topological invariant which can be calculated via $C=\frac{1}{2\pi }\int_{0}^{{2\pi }/{q}}{dk} \int_{0}^{2\pi }{d\delta }[\partial _{k}A_{\delta }-\partial _{\delta }A_{k}] $, where $A_{k}$ is the Berry connection defined by $A_{k}=i\langle \psi (k)|\partial_{k}|\psi (k)\rangle $. Similarly, we can define the Berry connection $A_{\delta }=i\langle \psi (k,\delta )|\partial _{\delta }|\psi (k,\delta
)\rangle $. 
To calculate it, we follow the method in Ref. \cite{doi:10.1143/JPSJ.74.1674} to directly perform the lattice computation of the Chern number. For the system with $\alpha =1/3$, we find that the Chern numbers in the three sub-band are $1$, $-2$ and $1$, respectively, for both the linear ($g=0$) and nonlinear ($g>0$) cases.

\subsection{Gap Solitons and Composition Relations}

\label{Gap Solitons Gap Solitons and Edge Soliton}

Besides the nonlinear Bloch waves, the nonlinear periodic system has another kind of solutions known as gap soliton solutions, which are spatially localized waves with the chemical potentials in the linear band gaps \cite{book}. It is found that the gap solitons and the nonlinear Wannier functions
match very well. The match
gets better as the periodic potential gets stronger \cite{PhysRevLett.102.093905, PhysRevA.80.063815}. The excellent match between the gap solitons and the nonlinear Wannier
functions suggests that the gap solitons be approximated by the the orbital wave functions of a unit cell since the the orbital wave functions can be taken as the Wannier functions when the periodic potential is stronger. As discussed in Refs. \cite{PhysRevLett.102.093905, PhysRevA.80.063815, PhysRevA.83.043610, 0953-4075-46-3-035301}, gap solitons develop in the linear band gaps and originate from the stable bound states of a single periodic well. So they can be divided different family according to the locations of the band gaps. On the other hand, the nonlinear Bloch band can be viewed as a lifted linear Bloch band by increasing the nonlinear interaction. However, the linear Bloch band can be viewed as an evolution from the discrete energy levels of an individual well. In particular, the gap solitons match the Wannier function well when the periodic potential is strong. Therefore, the gap solitons and nonlinear Bloch waves should share certain common features, which is called the `composition relation' \cite{Composition Relations}.
In this subsection, we shall explore the gap solitons in the superlattice system and their composition relations with the nonlinear Bloch waves.
\begin{figure}[tbp]
\includegraphics[width=9cm]{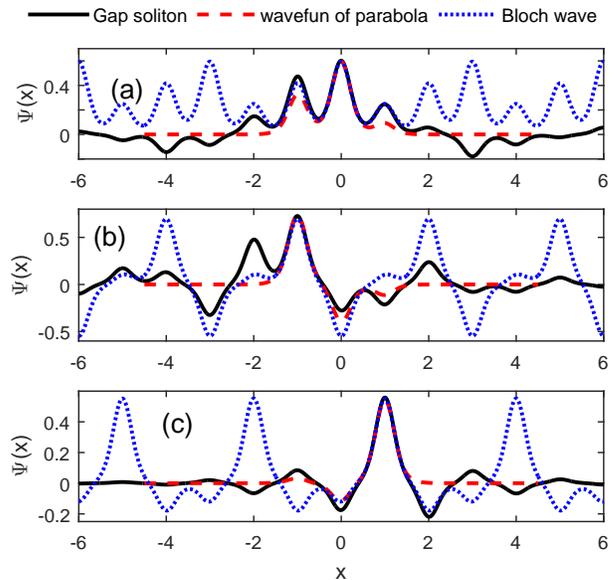}
\caption{The figure shows the gap solitons, the wave functions of parabola
and the nonlinear Bloch waves in the first and second band gaps. The
chemical potential of the gap solitons are used to be -5.1, -4.6 and -4.2.
All the nonlinear Bloch waves are taken at the center of the Brillouin zone (%
$k=0$).}
\label{Fig5}
\end{figure}

We solve the nonlinear Schr\"{o}dinger equation [Eq. (\ref{nonlinear Schrodinger equation})] directly to obtain the gap solitons shown in Fig. \ref{Fig5} (black solid lines). Under the parabolic approximation, we further solve the nonlinear Schr\"{o}dinger equation [Eq. (\ref{nonlinear Schrodinger equation})] to obtain the orbital wave functions of the corresponding parabola shown in Fig. \ref{Fig5} (red dashed lines). The chemical potentials are in three different band gaps from low to high. The states by two different methods coincide well. The good match indicates that they have the same origin. For the gap soliton in Fig. \ref{Fig5} (a), its chemical potential is in the first band gap. This state originates from the the ground state of the parabola. So the wave function has a main peak. However, the width of the parabola extends two period of $V_1(x)$. So the wave function has two extra little peaks.  The chemical potential of the gap soliton in Fig. \ref{Fig5} (b) is in the second band gap. This state originates from the the first excited state of the parabola. So one little peak in the wave function changes a sign.  The gap soliton in Fig. \ref{Fig5} (c) originates from the second excited state of the parabola and its chemical potential is in the second band gap. Both of the two little peaks in the wave function change the sign. Our results show the existence of a series of gap solitons which originate from the eigenstates of independent parabolas.


Our numerical results support
that the gap soliton are really fundamental and can be viewed
as building blocks for other stationary solutions of a nonlinear
periodic system, such as high-order gap solitons. Under the periodic boundary condition, we solve the nonlinear Bloch equation (\ref{Bloch equation}) to obtain the nonlinear Bloch waves shown in Fig. \ref{Fig5} by blue dotted lines. The
chemical potentials is set to be same as that of the corresponding gap solitons, and the wave vectors are are taken at the center of the Brillouin zone ($k=0$). Comparing the nonlinear Bloch waves
and the corresponding gap solitons  in Fig. \ref{Fig5}, we notice that the two waves
match very well within the single parabola. So a Bloch
wave at the center of the Brillouin zone can be viewed as a chain of
gap solitons pieced together. 

\subsection{Stability} \label{Stability}

In this subsection, we shall study the stability of the edge solitons against the
interaction strength following
the standard procedure \cite{Wu-Niu,Smerzi,Machholm}.
Since the unstable solution is sensitive to a small perturbation, we can
add a small perturbation $\Delta \Omega (x,t)$ to a known stationary
solution $\Omega (x)$ of the nonlinear Schr\"{o}dinger
equation (\ref{nonlinear Schrodinger equation})
\[
\Psi (x,t)=\left[ \Omega (x)+\Delta \Omega (x,t)\right] \exp (-i\mu t),
\]%
where $\Delta \Omega (x,t)=u(x)\exp (i\lambda t)+w^{\ast }(x)\exp (-i\lambda
^{\ast }t)$. 
Inserting the perturbation into time-dependent nonlinear Schr\"{o}dinger
equation and dropping the higher-order terms in ($u,v$), we then obtain
the linear eigenfunction
\begin{equation}
\left(
\begin{array}{cc}
\mathcal{L} & -g\Omega ^{2} \\
g\Omega ^{\ast 2} & -\mathcal{L}%
\end{array}%
\right) \left(
\begin{array}{c}
u \\
w%
\end{array}%
\right) =\lambda\left(
\begin{array}{c}
u \\
w%
\end{array}%
\right) ,  \label{LinearStability}
\end{equation}%
where $\mathcal{L}\equiv \frac{1}{2}\frac{d^{2}}{dx^{2}}-V\left( x\right)
-2g\left\vert \Omega \left( x\right) \right\vert ^{2}+\mu .$ Linear stability
of a soliton is determined by the energy spectrum of the linear
eigenfunction (\ref{LinearStability}). If all eigenvalues $\lambda$
are real, the solution of $\Omega (x)$ is stable. On the other hand, if there exists a finite imaginary part, the solution of $%
\Omega (x)$ would be unstable.

\begin{figure}[tbp]
\includegraphics[width=9cm]{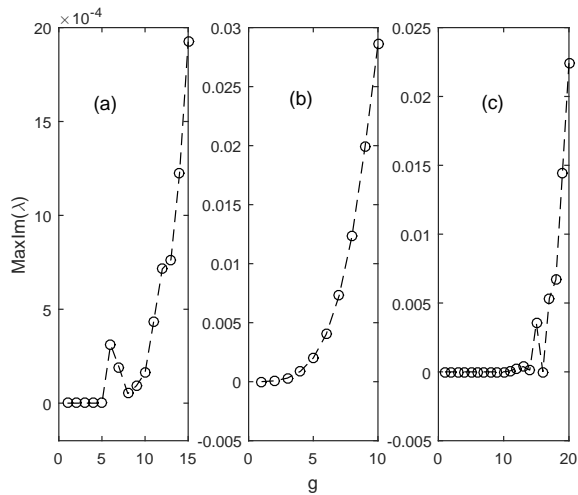}
\caption{The stability of the edge gap solitons: (a) $\protect\psi%
_{11}$, (b) $\protect\psi_{12}$ and (c) $\protect\psi_{23}$. }
\label{Fig6}
\end{figure}

The stabilities of the gap solitons have been discussed in Refs. \cite{PhysRevLett.102.093905, PhysRevA.80.063815, PhysRevA.83.043610, 0953-4075-46-3-035301} for several interacting periodic systems. Here, we focus our study on the stabilities of the edge solitons in bichromatic superlattices.
The stabilities of the edge gap solitons are displayed in Fig.~\ref{Fig6}. For the non-interaction case ($g=0$), the three edge gap solitons are reduced to the stationary solutions of linear Schr\"{o}dinger equation. When the interaction strength $g$ is increased, the three states will change from stable to unstable. The reason is that the confinement of the edge parabola is not strong enough to compensate the repulsive interaction and the kinetic energy. For the state $\Psi_{11}$ in Fig.~\ref{Fig6} (a), it is stable when $g<5$. However, $\Psi_{12}$ becomes unstable when $g>1$ in Fig.~\ref{Fig6} (b). This is due to the strong confinement of the left parabola to the particles which increases the kinetic energy and the repulsive interaction. For the edge gap soliton $\Psi_{23}$ in Fig.~\ref{Fig6} (c), it is still stable when $g=12$. The reason is that $\Psi_{23}$ originates from the first excited state of the right edge parabola. The half-width of the wave function is larger than the former states $\Psi_{11}$ and $\Psi_{12}$. So it has a low particle density which results the interactive energy is less than the former.

\section{Summary}

\label{Summary}

In summary, we explored nontrivial topological states in 1D nonlinear superlattice systems. Our study reveals that the nonlinear systems exhibit similar spectrum as the corresponding linear system and support the existence of topologically nontrivial edge gap solitons.  We unveiled the topological nature of the nonlinear Bloch bands by calculating the topological invariants of these bands. With the linear stability analysis, it is found that the edge gap solitons is stable when the nonlinear interaction is not strong enough. Our numerical results also verify that the composition relations between the gap solitons and nonlinear Bloch waves still hold true in the nonlinear superlattice systems. Our results will be helpful for understanding the effect of nonlinearity on topological states and exploring topologically nontrivial states in optical superlattice systems.

\begin{acknowledgments}
This work was supported by Hebei Provincial Natural Science Foundation of China
(Grant No. A2012203174, No. A2015203387, No. A2015203037) and
National Natural Science Foundation of China (NSFC) (Grant No. 10974169, No. 11475144 and No. 11304270). S C is supported by NSFC under Grants No. 11425419, No. 11374354 and No. 11174360, and the Strategic Priority Research Program (B) of the
Chinese Academy of Sciences  (No. XDB07020000).
\end{acknowledgments}


\end{document}